\documentclass[preprint,aps,prb]{revtex4}
\usepackage{epsfig}
\begin{document}

\title{Bounds on Localized Modes in the Crystal Impurity Problem}

\author{Harry J. Lipkin}\email{harry.lipkin@weizmann.ac.il}
\affiliation{Department of Particle Physics, Weizmann Institute of 
Science, Rehovot 76100, Israel}
\author{Philip D. Mannheim}\email{philip.mannheim@uconn.edu}
\affiliation{Department of Physics,
University of Connecticut, Storrs, CT 06269}

\date{February 7, 2006}

\begin{abstract}

Using general properties of the crystal site representation normal mode
matrix, we  provide some very simple bounds on localized modes in simple,
body-centered and face-centered cubic crystals with substitutional point
defects. We derive a trace condition constraint on the net change in
crystal eigenfrequencies caused by the introduction of a defect, with
the condition being a completely general one which holds for any
combination of central and non-central crystal force-constants and for
all-neighbor interactions. Using this condition we show that the
sufficient condition for producing localized modes in an arbitrary cubic
crystal by a mass change at the defect site is that the defect mass be
less than one half of that of the host atom mass which it replaces, and
that the sufficient condition for producing localized modes in an
arbitrary cubic crystal by force-constant changes alone is that the
defect site self force-constant be greater than twice that of the pure
crystal self force-constant of the host atom which it replaces.

\end{abstract}

\maketitle

\section{Introduction}

The substitutional insertion of a point defect impurity into an 
otherwise perfect host crystal will typically modify the spectrum of the
$3N$ normal modes of the crystal,\cite{Maradudin1965} leading in certain
circumstances (such as the insertion of impurities which are lighter in
mass than the host atoms which they replace or which are more strongly
coupled to the host crystal atoms than the ones they replace) to the
generation of modes with frequencies which lie beyond the band maximum
$\omega_{\rm max}$ of the crystal. Such modes will not be plane waves
which propagate throughout the crystal, but will instead fall off
exponentially fast away from the defect site and thus be localized to it.
Moreover, with the rest of the crystal atoms not participating appreciably
in such localized modes, the intensity of the defect in such modes will
be $N$ times larger than the intensity it would otherwise have had in a
crystal plane wave mode, to thus give the localized mode enough intensity
to render it observable. While such modes could be of relevance for
phenomena such as the $\rm{M\ddot{o}ssbauer}$ effect associated with the
insertion of $\rm{M\ddot{o}ssbauer}$ active defects into host crystals,
historical recoil-free fraction $\rm{M\ddot{o}ssbauer}$ studies only
involved an averaging over all the lattice modes of the system, to
thereby only allow one to infer the possible presence of localized modes
indirectly. However, with the advent of dedicated synchrotron rings it
became possible to monitor $\rm{M\ddot{o}ssbauer}$ active systems mode by
mode directly; and via nuclear resonant inelastic x-ray scattering
studies, modes lying beyond the host crystal phonon band maximum have now
explicitly been seen in the $^{57}{\rm Fe/Cu}$ 
and the $^{57}{\rm Fe/NiAl}$ systems.\cite{Seto2000,Parlinski2004}
Consequently, knowing whether the insertion of a defect into pure crystal
hosts might generate localized modes can be of great value for such
studies. 

In this paper we use a very straightforward trace technique to enable us
to identify some very general conditions under which such localized modes
can be produced. We shall restrict our study to the most straightforward
case of single point defects which are substitutionally inserted in the
three primitive simple, body-centered and face-centered cubic crystals
(these specific cases being amongst the most commonly experimentally
studied ones), though with our approach being quite generic, it could in
principle be adapted to encompass other crystal structures as well if
desired. We shall provide results for various combinations of mass and
force-constant changes, some which are specific to nearest-neighbor
force-constants and some of which are general to all-neighbor
force-constants. In Sec. II we derive the trace condition on which all of
our results are based, with the provided relation (Eq. (\ref{15}) below)
being an exact, all-neighbor relation which permits an arbitrary mass
change ($M$ replaced by $M^{\prime}$) at the defect site and arbitrary
force-constant changes between the defect atom and any other atom in the
entire crystal, despite which the resulting relation only involves the
self-force constant $A_{xx}(0,0)$ at the defect site and the change
$A^{\prime}_{xx}(0,0)$ in it. In Sec. III we apply our trace condition
to the case of an isotopic substitution where the only change is that in
the mass at the defect site, to show that while having a lighter impurity
($M^{\prime}<M$) is necessary for the generation of a localized mode, it
is the condition $M^{\prime}<M/2$ which is the sufficient one. In Secs.
IV and V we apply our trace condition to some straightforward
nearest-neighbor force-constant change cases for which there are
extremely simple exact analytic solutions to the impurity problem (some
typical examples of which being crystals with force-constants which are
central or which are isotropic), to find that for all of them the
sufficient condition for localized mode production by force-constant
changes alone is given as the requirement that
$A^{\prime}_{xx}(0,0)/A_{xx}(0,0)$ be greater than
$3/2$. Finally in Sec. V we examine more complicated force-constant
change cases and go beyond nearest-neighbor force-constants, and while
these situations do not admit of as straightforward a treatment as the
cases considered in Sec. IV, nonetheless for them we are still able to
extract a general sufficiency condition for the generation of localized
modes in an arbitrary crystal by force-constant changes alone, namely
that $A^{\prime}_{xx}(0,0)/A_{xx}(0,0)$ be greater than two.

\section{Derivation of the trace condition}

Since the very introduction of a defect breaks the translation invariance
of the lattice, to derive a condition such as one involving a trace, we
will need to work entirely in the coordinate space crystal site
representation. To actually derive the trace condition we recall that in
the harmonic approximation the equations of motion for the displacements
from equilibrium
$e^{-i\omega t}u_{\alpha}(\ell)$ of the atoms of a pure 3N-dimensional
cubic crystal lattice are given by
\begin{equation}
\sum_{\beta,\ell^{\prime}}\left[A_{\alpha \beta}(\ell, \ell^{\prime})
-w^2M(\ell^{\prime})\delta_{\alpha\beta}
\delta(\ell,\ell^{\prime})\right]u_{\beta}(\ell^{\prime})=0~~,
\label{1}
\end{equation}
where $\ell$ ranges from $0$ to $N-1$, $\alpha=x,y,z$, $M(\ell)$ 
is the mass of the atom at site $\ell$, and
$A_{\alpha\beta}(\ell,\ell^{\prime})$ are the second order
force-constants. Since Eq. (\ref{1}) is an eigenvalue equation for the
frequencies $\omega^2$, on defining a matrix $B_{\alpha
\beta}(\ell,\ell^{\prime})=
M(\ell)\delta_{\alpha\beta}\delta(\ell,\ell^{\prime})$, we
immediately see that the sum of the eigenfrequencies of the crystal is
given by
\begin{equation}
\sum_{i=1}^{3N} \omega_i^2={\rm Tr}AB^{-1}~~.
\label{2}
\end{equation}
For a pure crystal in which all masses are equal to a common
$M(\ell)=M$ and all self force-constants are equal to a common 
$A_{xx}(\ell,\ell)=A_{yy}(\ell,\ell)=A_{zz}(\ell,\ell)=A_{xx}(0,0)$, Eq.
(\ref{2}) reduces to
\begin{equation}
\frac{1}{3N}\sum_{i=1}^{3N} \omega_i^2=\frac{A_{xx}(0,0)}{M}~~.
\label{3}
\end{equation}
With the left-hand side of Eq. (\ref{3}) being recognized as $\mu_2$, the
second moment of the density of states, Eq. (\ref{3}) thus recovers the
well-known relation for pure crystals 
\cite{Mannheim1972}
\begin{equation}
\mu_2=\frac{A_{xx}(0,0)}{M}~~.
\label{4}
\end{equation}

For the system with a substitutional point impurity of mass
$M^{\prime}$ located at the origin of coordinates and changed
force-constants $A^{\prime}_{\alpha\beta}(\ell,\ell^{\prime})$, the
displacements from equilibrium are now given as
$e^{-i\omega^{\prime} t}u_{\alpha}(\ell)$,  with Eq. (\ref{1}) then
being replaced by
\begin{equation}
\sum_{\beta,\ell^{\prime}}\left[A_{\alpha \beta}(\ell, \ell^{\prime})
-w^{\prime 2}M\delta_{\alpha\beta}
\delta(\ell,\ell^{\prime})\right]u_{\beta}(\ell^{\prime})
=\sum_{\beta,\ell^{\prime}}V_{\alpha \beta}(\ell,
\ell^{\prime})u_{\beta}(\ell^{\prime})~~,
\label{5}
\end{equation}
where the changes from the pure crystal case are described by the
perturbation
\begin{equation}
V_{\alpha \beta}(\ell, \ell^{\prime})
=-w^{\prime 2}(M-M^{\prime})\delta_{\alpha\beta}
\delta(\ell,0)\delta(\ell^{\prime},0)+A_{\alpha \beta}(\ell, 
\ell^{\prime})-
A^{\prime}_{\alpha \beta}(\ell, \ell^{\prime})~~.
\label{6}
\end{equation}
On now defining a matrix $B^{\prime}_{\alpha
\beta}(\ell,\ell^{\prime})=
M\delta_{\alpha\beta}\delta(\ell,\ell^{\prime})+
(M^{\prime}-M)\delta_{\alpha\beta}
\delta(\ell,0)\delta(\ell^{\prime},0)$,
the sum of the eigenfrequencies of the perturbed crystal is then given by
\begin{equation}
\sum_{i=1}^{3N} \omega^{\prime 2}_i={\rm Tr}
A^{\prime}[B^{\prime}]^{-1}~~,
\label{7}
\end{equation}
with the change in the sum of the eigenfrequencies thus being given by
\begin{equation}
\sum_{i=1}^{3N} [\omega^{\prime 2}_i-\omega_i^2]={\rm Tr}
A^{\prime}[B^{\prime}]^{-1}-{\rm Tr}
AB^{-1}~~.
\label{8}
\end{equation}
The utility of Eq. (\ref{8}) is that its left-hand side measures whether
modes are shifted to higher or lower frequency, while on its right-hand
side it is only the sites which are explicitly involved in $V_{\alpha
\beta}(\ell, \ell^{\prime})$ which do not drop out of the
difference between the two 3N-dimensional traces. Our task is thus to
first evaluate the right-hand side of Eq. (\ref{8}) and then to seek
constraints on its left-hand side.

In the simplest case where the only change is a mass change at the defect
site, it is only the defect site contribution itself which is not
cancelled in the trace difference, with the full change in frequency
immediately being given by
\begin{equation}
\sum_{i=1}^{3N} [\omega^{\prime 2}_i-\omega_i^2]=
\frac{3A_{xx}(0,0)}{M^{\prime}}-
\frac{3A_{xx}(0,0)}{M}~~.
\label{9}
\end{equation}
When there is also a change in force-constant at the defect site, we
recall that by Newton's third law of motion there must also be changes in
the force-constants at other sites too. Moreover, this same Newtonian law
entails that the force-constants of a perturbed system with a defect
have to obey
\begin{equation}
A^{\prime}_{\alpha\beta}(\ell,\ell)=-\sum_{\ell^{\prime} \neq
\ell}A^{\prime}_{\alpha\beta}(\ell^{\prime},\ell)
\label{10}
\end{equation}
for all $\ell$ in exactly the same way as the pure crystal
force-constants have to obey
\begin{equation}
A_{\alpha\beta}(\ell,\ell)=-\sum_{\ell^{\prime} \neq
\ell}A_{\alpha\beta}(\ell^{\prime},\ell)~~.
\label{11}
\end{equation}
With the $x$, $y$ and $z$ directions being equivalent in cubic
crystals, in the general case which allows for arbitrary force-constant
changes, Eq. (\ref{8}) takes the form 
\begin{eqnarray}
\frac{1}{3}\sum_{i=1}^{3N} [\omega^{\prime 2}_i-\omega_i^2]=&& 
\frac{A^{\prime}_{xx}(0,0)}{M^{\prime}}-\frac{A_{xx}(0,0)}{M}
\nonumber \\
&&+\frac{A^{\prime}_{xx}(1,1)}{M}-\frac{A_{xx}(1,1)}{M}
+\frac{A^{\prime}_{xx}(2,2)}{M}-\frac{A_{xx}(2,2)}{M}+....~~.
\label{12}
\end{eqnarray}
As regards the pure and the perturbed force-constants, we recall that in
terms of the two-body interatomic potential $\phi(r)$, the force-constants
between an atom vibrating around site $R_{\alpha}(\ell)$ and one vibrating
around the origin are defined as
\begin{equation}
A_{\alpha\beta}(\ell,0)=-
\left[\frac{\partial^2\phi(r)}{\partial
u_{\alpha}(\ell)\partial u_{\beta}(\ell)}\right]\bigg{|}_0=
-\left[\frac{\phi^{\prime\prime}(r)}{r^2}
-\frac{\phi^{\prime}(r)}{r^3}\right]R_{\alpha}(\ell)R_{\beta}(\ell)
-\frac{\phi^{\prime}(r)}{r}\delta_{\alpha\beta}~~,
\label{13}
\end{equation}
as calculated at the equilibrium separation $r$. 
Then, since the effect of the introduction of the defect is to alter the
two-body potential between the defect and the host atoms while not
affecting the potentials between any two host atoms themselves, the
only non-self force-constants which will change will then be the
$A^{\prime}_{\alpha\beta}(\ell,0)$ with
$\ell \neq 0$, with Eq. (\ref{10}) then obliging the self
force-constants at the defect site and those at the $\ell \neq 0$ sites to
respectively change as
\begin{eqnarray}
A^{\prime}_{\alpha\beta}(0,0)&=&-\sum_{\ell \neq
0}A^{\prime}_{\alpha\beta}(\ell,0)~~,~~
\nonumber \\
A^{\prime}_{\alpha\beta}(\ell,\ell)&=&-A^{\prime}_{\alpha\beta}(\ell,0)
-\sum_{\ell^{\prime} \neq \ell,0}A_{\alpha\beta}(\ell,\ell^{\prime})
=-A^{\prime}_{\alpha\beta}(\ell,0)
+A_{\alpha\beta}(\ell,\ell)+A_{\alpha\beta}(\ell,0)~~.
\label{14}
\end{eqnarray}
Given these relations and Eqs. (\ref{10}) and (\ref{11}), Eq.
(\ref{12}) can thus be simplified to
\begin{eqnarray}
\frac{1}{3}\sum_{i=1}^{3N} [\omega^{\prime 2}_i-\omega_i^2]&&= 
\frac{A^{\prime}_{xx}(0,0)}{M^{\prime}}-\frac{A_{xx}(0,0)}{M}
-\frac{1}{M}\sum_{\ell \neq 0}\left[A^{\prime}_{xx}(\ell,0)
-A_{xx}(\ell,0)\right]
\nonumber \\
&&=\frac{A^{\prime}_{xx}(0,0)}{M^{\prime}}+\frac{A^{\prime}_{xx}(0,0)}{M}
-\frac{2A_{xx}(0,0)}{M}~~.
\label{15}
\end{eqnarray}
Equation (\ref{15}) is our main result, and is derived here with no
restriction at all on the number of neighbors of the defect for which
the force-constants might change. Nor does it presuppose any relation
between $\phi^{\prime\prime}(r)$ and $\phi^{\prime}(r)/r$. Despite the
fact that Eq. (\ref{15}) conveniently only involves the self
force-constant at the defect site, nonetheless it is an all-neighbor
result, one which additionally holds for any combination of central
and non-central force-constants.\cite{footnote1}

\section{Application to the mass defect case}

In the treatment of crystal impurity problem it is conventional to solve
Eq. (\ref{5}) by the lattice Green's function technique, and since we will
use some of its aspects to constrain the left-hand side of Eq. (\ref{15}),
we briefly recall the procedure. One first introduces the dynamical matrix
of the pure crystal
\begin{equation}
D_{\alpha\beta}(\vec{\bf
k})=\frac{1}{M}\sum_{\ell}A_{\alpha\beta}(0,\ell)e^{-i\vec{\bf k} \cdot
\vec{\bf R}(\ell)}
\label{16}
\end{equation}
as expressed in terms of the phonon modes $\vec{\bf k}$ of the
translational invariant pure crystal, and then defines its eigenvectors
and eigenvalues according to
\begin{eqnarray}
\sum_{\beta}D_{\alpha\beta}(\vec{\bf
k})\sigma_{\beta}^{j}(\vec{\bf
k})&=&\omega_{j}^2(\vec{\bf k})\sigma_{\alpha}^{j}(\vec{\bf k})~~,
\nonumber \\
\sum_{\alpha}\sigma_{\alpha}^{*j}(\vec{\bf
k})\sigma_{\alpha}^{j^{\prime}}(\vec{\bf k})&=&\delta_{jj^{\prime}}~~,~~
\sum_{j}\sigma_{\alpha}^{*j}(\vec{\bf
k})\sigma_{\beta}^{j}(\vec{\bf k})=\delta_{\alpha\beta}~~,
\label{17}
\end{eqnarray}
and then uses these eigenvectors and eigenvalues to construct the pure
crystal lattice Green's functions according to
\begin{equation}
g_{\alpha\beta}(\omega;\ell,\ell^{\prime})=
\frac{1}{NM}\sum_{\vec{\bf k},j}\frac{\sigma_{\alpha}^{*j}
(\vec{\bf k})\sigma_{\beta}^{j}(\vec{\bf k})
e^{i\vec{\bf k}\cdot[\vec{\bf R}(\ell^{\prime})-\vec{\bf
R}(\ell)]}}{[\omega_{j}^2(\vec{\bf k})-\omega^2]}
\label{18}
\end{equation}
as summed over the polarizations $j=(1,2,3)$ and momenta $\vec{\bf k}$ of
all the modes in the Brillouin zone. As constructed these Green's
functions obey
\begin{equation}
\sum_{\ell,\beta}A_{\alpha\beta}(0,\ell)g_{\alpha^{\prime}\beta}
(\omega;\ell,\ell^{\prime}))=
M\omega^2g_{\alpha^{\prime}\alpha}(\omega;0,\ell^{\prime})
+\frac{\delta_{\alpha,\alpha^{\prime}}}{N}\sum_{\vec{\bf k}}
e^{i\vec{\bf k}\cdot\vec{\bf R}(\ell^{\prime})}~~,
\label{19}
\end{equation}
and thus immediately allow us to solve Eq. (\ref{5}) in the form
\begin{equation}
u_{\alpha}(\ell)=\sum_{\ell^{\prime},\ell^{\prime\prime},
\beta,\gamma}g_{\alpha\beta} (\omega^{\prime};\ell,\ell^{\prime})
V_{\beta\gamma}(\ell^{\prime},\ell^{\prime\prime})
u_{\gamma}(\ell^{\prime\prime})~~,
\label{20}
\end{equation}
with the eigenmodes being given as the solutions to the
($3N$-dimensional) determinantal condition $|1-G_0V|=0$ as written in an
obvious notation.

For the simplest case of just a change in mass at the defect site, Eq.
(\ref{20}) requires the three components of the defect displacement vector
to obey
\begin{eqnarray}
u_{x}(0)&=&-(M-M^{\prime})\omega^{\prime 2}g_{xx}
(\omega^{\prime};0,0)u_{x}(0)~~,
\nonumber\\
u_{y}(0)&=&-(M-M^{\prime})\omega^{\prime 2}g_{yy}
(\omega^{\prime};0,0)u_{y}(0)~~,
\nonumber \\
u_{z}(0)&=&-(M-M^{\prime})\omega^{\prime 2}g_{zz}
(\omega^{\prime};0,0)u_{z}(0)~~,
\label{21}
\end{eqnarray}
where the cubic symmetry of the host lattice requires that $g_{xx}
(\omega^{\prime};0,0)$, $g_{yy} (\omega^{\prime};0,0)$ and $g_{zz}
(\omega^{\prime};0,0)$ all be equal to each other, and thus that each
one of them be given as
\begin{eqnarray}
g_{xx} (\omega^{\prime};0,0)&=&\frac{1}{NM}\sum_{\bar{k},j}
\frac{\sigma_x^{*j}(\vec{\bf k})\sigma_x^{j}(\vec{\bf k}) }
{[\omega_j^2(\vec{\bf k})-\omega^{\prime 2}]}
=\frac{1}{3}\left[g_{xx}(\omega^{\prime};0,0) 
+g_{yy}(\omega^{\prime};0,0)  +g_{zz}(\omega^{\prime};0,0)\right]
\nonumber \\
&=&
\frac{1}{3NM}\sum_{\bar{k},j}
\frac{[\sigma_x^{*j}(\vec{\bf k})\sigma_x^{j}(\vec{\bf k}) 
+\sigma_y^{*j}(\vec{\bf k})\sigma_y^{j}(\vec{\bf k}) 
+\sigma_z^{*j}(\vec{\bf k})\sigma_z^{j}(\vec{\bf k})]}
{[\omega_j^2(\vec{\bf k})-\omega^{\prime 2}]}
\nonumber \\
&=&
\frac{1}{3NM}\sum_{\bar{k},j}
\frac{1}
{[\omega_j^2(\vec{\bf k})-\omega^{\prime 2}]}
\nonumber \\
&=&\frac{1}{NM}\sum_{i=1}^N
\frac{1}{(\omega_i^2-\omega^{\prime 2})}
=\frac{1}{M}\int_0^{\omega^2_{\rm
max}}d\omega^{2}\frac{\nu(\omega^{2})}{(\omega^2-\omega^{\prime
2})}
\label{22}
\end{eqnarray}
as now summed over $N$ threefold degenerate pure crystal 
eigenmodes $\omega_i^2$.\cite{footnote2} And with the determinantal
condition then being given by 
$|1-G_0V|=(M^{\prime}/M)^3[1-\rho(\omega^{\prime 2})S(\omega^{\prime
2})]^3(1)^{3N-3}=0$, in the mass defect case the perturbed
simple, body-centered and face-centered cubic crystal modes are thus given
as the solutions to the familiar
\cite{Maradudin1965}
\begin{equation}
1-\rho(\omega^{\prime 2})S(\omega^{\prime 2})=0~~,
\label{23}
\end{equation}
where $\rho(\omega^{\prime 2})$ is given by
\begin{equation}
\rho(\omega^{\prime 2})=\frac{M}{M^{\prime}}-1
\label{24}
\end{equation}
and $S(\omega^{\prime 2})$ is given by 
\begin{equation}
S(\omega^{\prime 2})=-1-M\omega^{\prime
2}g_{xx}(\omega^{\prime };0,0)=\frac{1}{N}
\sum_{i=1}^{N}\frac{\omega_i^2}{(\omega^{\prime 2}-\omega_i^2)}~~,  
\label{25}
\end{equation}
as summed over the $N$ eigenfrequencies $\omega_i^2$ of the pure
crystal.

\begin{figure}
\centerline{\epsfig{file=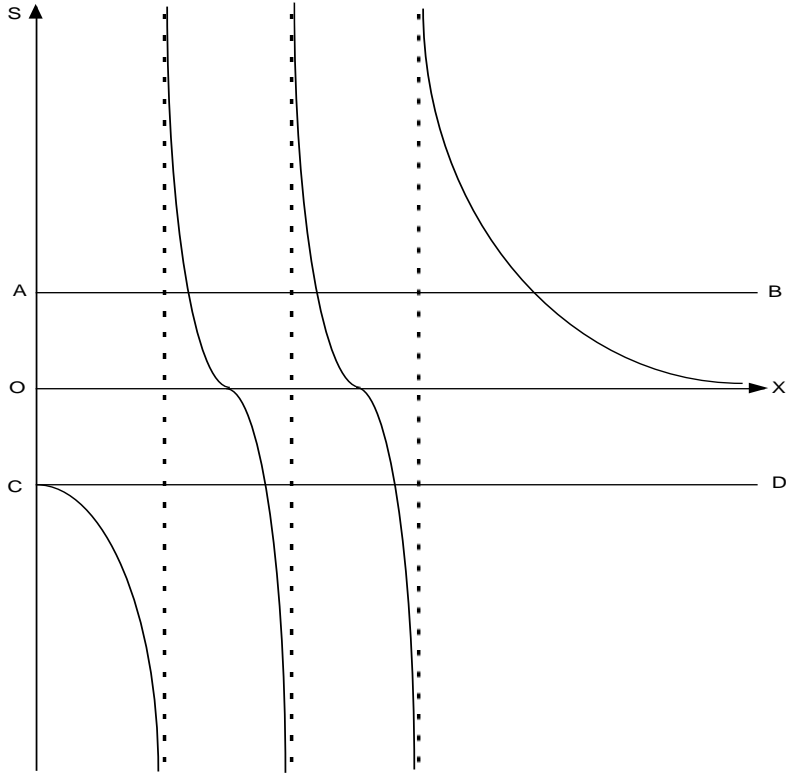,width=5.1in,height=5.1in}}
\medskip
\caption{Plot of $S(\omega^{\prime 2})$ versus $\omega^{\prime 2}$. The
line $AB$ corresponds to the value of $1/\rho(\omega^{\prime 2})$ in a
typical lighter defect case with $M^{\prime}<M$, while the line $CD$
corresponds to the $1/\rho(\omega^{\prime 2})$ associated with
an infinitely heavy defect case.}
\label{Fig. (1)}
\end{figure}

For our purposes here we note that the function $S(\omega^{\prime 2})$
is divergent at every pure crystal frequency, to
thus contain the set of $N$ asymptotes indicated in Fig. (1). 
$S(\omega^{\prime 2})$ starts out equal to minus one at
$\omega^{\prime 2}=0$, and decreases to minus infinity at the first
asymptote, with
$S(\omega^{\prime 2})$ thus having no zero below the lowest pure crystal
band mode. Since the derivative of
$S(\omega^{\prime 2})$, viz. 
\begin{equation}
\frac{dS(\omega^{\prime 2})}{d\omega^{\prime 2}}=-\frac{1}{N}
\sum_{i=1}^{N}\frac{\omega_i^2}{(\omega^{\prime 2}-\omega_i^2)^2}   
\label{26}
\end{equation}
is negative definite, the function $S(\omega^{\prime 2})$ falls
monotonically between asymptotes, to thus have a zero somewhere on the
$OX$ axis between any two adjacent pure crystal modes, while beyond the
last pure crystal frequency eigenmode at $\omega^2_{\rm max}$ the
function $S(\omega^{\prime 2})$ is positive, falling to zero at
$\omega^{\prime 2}=\infty$. As introduced, the function $S(\omega^{\prime
2})$ thus has $N-1$ zeroes inside the pure crystal phonon band and one 
zero outside.

In consequence of this structure for $S(\omega^{\prime 2})$, we see that
the curve $1/\rho(\omega^{\prime 2})=M^{\prime}/(M-M^{\prime})$ will
intercept the
$S(\omega^{\prime 2})$ curve $N$ times no matter what the value of
$M^{\prime}/M$,\cite{footnote3} with Eq. (\ref{21}) thus always having
the same number of mode solutions as the pure crystal (viz. $N$
solutions for each of the three $u_x(0)$, $u_y(0)$ and $u_z(0)$ sectors of
Eq. (\ref{21})), with Eq. (\ref{23}) and the threefold degeneracy of
Eq. (\ref{21}) thus accounting for the complete spectrum of
eigenfrequencies of the perturbed crystal. When $M^{\prime}$ is less than
$M$ (typical intercept $AB$) the $N$ modes which satisfy Eq. (\ref{23})
are all shifted to higher frequency compared to their pure crystal
counterparts, with both sides of Eq. (\ref{9}) thus consistently being
positive. Similarly, when $M^{\prime}$ is greater than $M$ (typical
intercept $CD$) the $N$ modes which satisfy Eq. (\ref{23}) are all
shifted to lower frequency compared to their pure crystal counterparts,
with both sides of Eq. (\ref{9}) then consistently being negative.
Finally, once the eigenfrequencies have been found, a return to Eq.
(\ref{5}) will then allow a determination of all of the $3N$
displacements $u_{\alpha}(\ell)$ in every mode, to then completely
specify the crystal displacements of the perturbed crystal.

Since the zeroes of $S(\omega^{\prime 2})$ other than the final one at
$\omega^{\prime 2}=\infty$ lie between adjacent asymptotes, an initial
glance at Fig. (1) would suggest that when shifted upwards, each of
$(N-1)$ modes other than the mode at the band maximum itself could be
shifted upward as far as the immediate next asymptote, to suggest
that the quantity $\sum_{i=1}^{N-1} [\omega^{\prime 2}_i-\omega_i^2]$
could be as large as $\omega_{\rm max}^2-\omega_1^2 \sim \omega_{\rm
max}^2$ (the net shift between the highest and lowest in-band frequency
modes). However, we will now show that in fact the net shift must be much
less than this, an outcome that will sharply constrain the implications
of Eq. (\ref{15}). To this end we need to obtain a bound on the sum of the
in-band zeroes, $x^2_i$, of $S(\omega^{\prime 2})$, and note that since
the quantity $S(\omega^{\prime 2})\Pi_i(\omega^{\prime 2}-\omega_i^2)$ is
an $N-1$ dimensional polynomial in $\omega^{\prime 2}$, the sum of its
zeroes can immediately be given as
\begin{equation}
\sum_{i=1}^{N-1}x_i^2 =\sum_{i=1}^{N}\omega^{2}_i 
-\frac{\sum_{i=1}^{N}\omega^{4}_i}
{\sum_{i=1}^{N}\omega^{2}_i}=
\sum_{i=1}^{N}\omega^{2}_i 
-\frac{\mu_4}
{\mu_2}~~, 
\label{27}
\end{equation}
where
\begin{equation}
\mu_{n}=\frac{1}{N}\sum_{i=1}^{N}\omega_i^{n}~~.
\label{28}
\end{equation}
On defining the positive semi-definite quantity $\alpha$ via
\begin{equation}
\mu_4-\mu_2^2=\langle \omega^4\rangle-\langle \omega^2\rangle^2 =
\langle [\omega^2-\langle \omega^2\rangle]^2\rangle =\alpha~~,
\label{29}
\end{equation}
through use of Eq. (\ref{3}) we then obtain 
\begin{equation}
\sum_{i=1}^{N-1}[x^{2}_i -\omega^{2}_{i}] 
= \omega_{\rm max}^2-\mu_2 -\frac{\alpha}{\mu_2}
= \omega_{\rm max}^2-\frac{A_{xx}(0,0)}{M}-\frac{\alpha}{\mu_2}~~.  
\label{30}
\end{equation}
Now in general it can be shown that the quantity
$A_{xx}(0,0)/M$ is related to $\omega^2_{\rm max}$
according to\cite{Mannheim1972}
\begin{equation}
\omega_{\rm max}^2=\frac{2A_{xx}(0,0)(1+Z)}{M}  
\label{31}
\end{equation}
where the quantity $Z$ is given by 
\begin{equation}
Z=\frac{\sum_{\ell}A_{xx}(0,\ell)}{A_{xx}(0,0)} 
\label{32}
\end{equation}
as summed over the even neighbors of the $\ell=0$ site alone. With
this sum only beginning with the second nearest neighbors, to good
approximation we can neglect the contribution of $Z$, and can thus
rewrite Eq. (\ref{30}) as 
\begin{equation}
\sum_{i=1}^{N-1}[x^{2}_i -\omega^{2}_{i}] 
= \frac{\omega_{\rm max}^2}{2}-\frac{\alpha}{\mu_2}~~.  
\label{33}
\end{equation}
With $\alpha$ being positive, we thus see that the quantity
$\sum_{i=1}^{N-1} [x^{2}_i-\omega_i^2]$ cannot be larger than
$\omega_{\rm max}^2/2$, with the zeroes of $S(\omega^{\prime 2})$ on
average being no more than midway between adjacent asymptotes rather than
close to the immediately adjacent higher ones.  

Further support for this result can be obtained by considering
Eq. (\ref{23}) in the infinitely heavy defect limit in which
$M^{\prime}=\infty$, $\rho(\omega^{\prime 2})=-1$, corresponding to the
line $CD$ in Fig. (1). In this case Eq. (\ref{9}) reduces to
\begin{equation}
\sum_{i=1}^{N} [\omega^{\prime 2}_i-\omega_i^2]=
-\frac{A_{xx}(0,0)}{M}=-\frac{\omega_{\rm max}^2}{2}~~,
\label{34}
\end{equation}
with the structure of Fig. (1) then yielding
\begin{equation}
-\frac{\omega_{\rm max}^2}{2} 
= 0+(\omega_2^{\prime 2}-\omega_1^2)+(\omega_3^{\prime
2}-\omega_2^2)+...-\omega_{\rm max}^2
>\sum_{i=1}^{N-1}
[x^{2}_i-\omega_i^2]-\omega_{\rm max}^2~~,
\label{35}
\end{equation}
from which the bound
\begin{equation}
\sum_{i=1}^{N-1}
[x^{2}_i-\omega_i^2]<\frac{\omega_{\rm max}^2}{2}
\label{36}
\end{equation}
then follows.

Having now established the bound on the zeroes of $S(\omega^{\prime 2})$
which is given in Eqs. (\ref{33}) and (\ref{36}), we now note that for the
pure mass defect case Eq. (\ref{9}) requires that 
\begin{equation}
\sum_{i=1}^{N-1}[\omega^{\prime 2}_i
-\omega^{2}_{i}]+\omega^{\prime 2}_{\rm max} -\omega^{ 2}_{\rm
max}=\frac{A_{xx}(0,0)}{M^{\prime}}-
\frac{A_{xx}(0,0)}{M}~~,
\label{37}
\end{equation}
where $\omega^{\prime 2}_{\rm max}$ is the largest shifted frequency.
Since in the $M^{\prime}<M$ case the in-band modes have to lie to the
left of the $x_i^2$ zeroes of $S(\omega^{\prime 2})$, the
in-band $\sum_{i=1}^{N-1}[\omega^{\prime 2}_i -\omega^{2}_{i}]$ can
then never be any larger than $\omega_{\rm max}^2/2\sim A_{xx}(0,0)/M$.
Consequently, no matter what the value of $\alpha$ of Eq. (\ref{29}) (a
quantity which varies from one crystal to the next), and no matter which
particular crystal one might choose, if $M^{\prime}$ is less than $M/2$,
the largest perturbed eigenfrequency $\omega^{\prime 2}_{\rm max}$
would than have to be larger than the pure crystal maximum $\omega^{
2}_{\rm max}$, and  not only would it then lie beyond the pure crystal
phonon band, once $M^{\prime}$ has been reduced enough to move the
in-band modes as far to higher frequency as they are able to go (viz.
to the zeroes of $S(\omega^{\prime 2})$), further reduction in
$M^{\prime}$ would then cause $\omega^{\prime 2}_{\rm max}$ to move 
further and further away from the band maximum. Hence while the condition
$M^{\prime}<M$ is a necessary condition for producing a mode beyond the
band maximum in simple, body-centered and face-centered cubic crystals,
it is $M^{\prime}<M/2$ which is the sufficient one, a result previously 
obtained\cite{Mannheim1972} by entirely different means.\cite{footnote 4}

\section{The force-constant change case}

Because of Newton's third law of motion, in the presence of
force-constant changes the matrix $V_{\alpha\beta}(\ell,\ell^{\prime})$ of
Eq. (\ref{6})  will necessarily involve more atoms than just the one
at the defect site and the problem is essentially intractable other than
numerically if $V_{\alpha\beta}(\ell,\ell^{\prime})$ extends beyond the
nearest neighbors of the defect. Moreover, even in the event that one
restricts the force-constant changes to the defect and its nearest
neighbors alone, for the simple cubic, body-centered cubic and
face-centered cubic crystals the defect and its nearest neighbors
respectively consist of a cluster of seven, nine and thirteen atoms,
making $V_{\alpha\beta}(\ell,\ell^{\prime})$ 21-, 27- and 39-dimensional
in those respective cases. Fortunately, because of the cubic crystal
$O_h$ symmetry at the defect site the
$V_{\alpha\beta}(\ell,\ell^{\prime})$ matrix can be block diagonalized in
the irreducible representations of the octahedral group, with typical
decompositions\cite{Maradudin1965}
\begin{eqnarray}
\Gamma_{\rm sc}&=&A_{1g}+E_g+F_{1g}+
F_{2g}+3F_{1u}+F_{2u}
\nonumber \\
\Gamma_{\rm bcc}&=&A_{1g}+E_g+F_{1g}+
2F_{2g}+A_{2u}+E_u+3F_{1u}+F_{2u}
\nonumber \\
\Gamma_{\rm fcc}&=&A_{1g}+A_{2g}+2E_g+2F_{1g}+
2F_{2g}+A_{2u}+E_u+4F_{1u}+2F_{2u}~~.
\label{38}
\end{eqnarray}
Since the displacement of the defect atom itself transforms as a
3-dimensional vector, the defect displacements must be located entirely in
the $F_{1u}$ modes, with all of the other relevant $O_h$ irreducible
representations being built solely out of appropriate
\cite{Maradudin1965} linear combinations of the displacements of the
nearest neighbors of the defect. As such the defect displacement appears
in 3-dimensional, 3-dimensional and 4-dimensional representations in the
three respective cubic crystal cases, with the defect sector
thus leading to respective 3-dimensional, 3-dimensional and 4-dimensional
blocks in Eq. (\ref{20}) each one of which (just as in Eq. (\ref{21})) is
threefold degenerate. While these blocks are still quite
complicated,\cite{footnote5} in the case of central force-constants alone
(as well as in some other specific cases such as isotropic force-constant
crystals which we discuss in Sec. V) the determinantal condition
$|1-G_0V|=0$ associated with the
$F_{1u}$ block can be treated completely analytically, with both the
body-centered cubic and face-centered cubic crystal eigenmodes being found
\cite{Mannheim1968,Mannheim1971} to obey precisely the same Eq.
(\ref{23}) as before, viz.
\begin{equation}
1-\rho(\omega^{\prime 2})S(\omega^{\prime 2})=0~~,
\label{39}
\end{equation}
save only that this time $\rho(\omega^{\prime 2})$ is given by
\begin{equation}
\rho(\omega^{\prime 2})=\frac{M}{M^{\prime}}-1 +\frac{2\omega^{\prime
2}}{\omega_{\rm
max}^2}\left[1-\frac{A_{xx}(0,0)}{A^{\prime}_{xx}(0,0)}\right]~~.
\label{40}
\end{equation}

\begin{figure}
\centerline{\epsfig{file=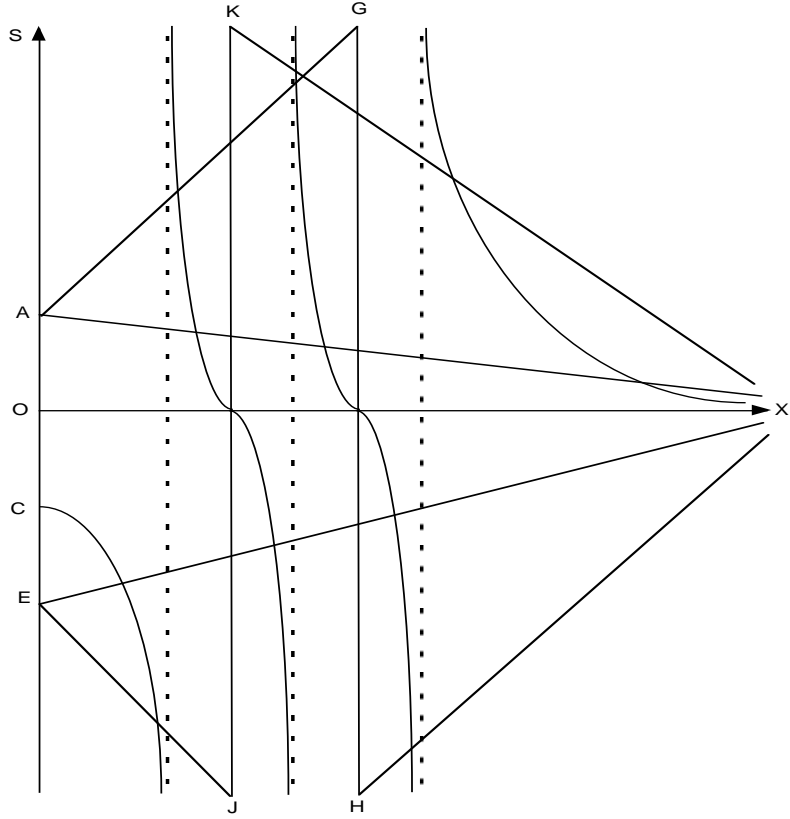,width=5.1in,height=5.1in}}
\medskip
\caption{Plot of $S(\omega^{\prime 2})$ versus $\omega^{\prime
2}$. The line $AX$ corresponds to the value of $1/\rho(\omega^{\prime 2})$
associated with a defect with $M^{\prime}<M$ and
$A_{xx}^{\prime}(0,0)>A_{xx}(0,0)$, the
line $EX$ corresponds to the $1/\rho(\omega^{\prime 2})$ associated
with
$M^{\prime}>M$ and
$A_{xx}^{\prime}(0,0)<A_{xx}(0,0)$, the
curve $AGHX$ corresponds corresponds to the $1/\rho(\omega^{\prime 2})$
associated with $M^{\prime}<M$ and
$A_{xx}^{\prime}(0,0)<A_{xx}(0,0)$, and the
curve $EJKX$ corresponds corresponds to the $1/\rho(\omega^{\prime 2})$
associated with $M^{\prime}>M$ and
$A_{xx}^{\prime}(0,0)>A_{xx}(0,0)$.}
\label{Fig. (2)}
\end{figure}

For our purposes here we note that since the function $S(\omega^{\prime
2})$ is the same one as discussed previously, it still has the
asymptote structure shown in Fig. (1). To determine the number of times
the function $1/\rho(\omega^{\prime 2})$ will intersect $S(\omega^{\prime
2})$ requires considering each possible combination of mass and
force-constant changes in Eq. (\ref{40}) separately. For $M^{\prime}\leq
M$ and $A_{xx}^{\prime}(0,0)>A_{xx}(0,0)$ the quantity
$1/\rho(\omega^{\prime 2})$ is everywhere positive yielding the typical
curve $AX$ in Fig. (2) with its $N$ intercepts (the point $A$ is at
$+\infty$ when $M^{\prime}=M$), with each crystal eigenmode having been
shifted upwards to a higher frequency. For $M^{\prime}\geq M$ and
$A_{xx}^{\prime}(0,0)<A_{xx}(0,0)$ the quantity $1/\rho(\omega^{\prime
2})$ is everywhere less than or equal to $-1$ yielding the typical curve
$EX$ in Fig. (2) with its $N$ intercepts (the point $E$ is at
$-\infty$ when $M^{\prime}=M$), with each crystal eigenmode having been
shifted downwards. For $M^{\prime}<M$ but
$A_{xx}^{\prime}(0,0)<A_{xx}(0,0)$ the quantity $1/\rho(\omega^{\prime
2})$ has to diverge somewhere, yielding the typical curve $AGHX$ in Fig.
(2) with its $N$ intercepts if the divergence in $1/\rho(\omega^{\prime
2})$ falls inside the band,\cite{footnote6} and analogously also yielding
$N$ intercepts if the divergence is outside the band (not
shown).\cite{footnote7} Finally, for
$M^{\prime}>M$ and $A_{xx}^{\prime}(0,0)>A_{xx}(0,0)$ there will again
be $N$ intercepts (typical curve $EJKX$).\cite{footnote8} 

As we thus see, no matter what particular values $M^{\prime}/M$ and
$A_{xx}^{\prime}(0,0)/A_{xx}(0,0)$ might actually take, in all cases there
are precisely $N$ intercepts, and thus precisely $3N$ eigenmodes in the
$F_{1u}$ sector. However, since this exhausts the number of degrees of
freedom for the problem, Eq. (\ref{20}) cannot generate any further
eigenmode solutions.  Consequently, none of the determinantal conditions
associated with any of the other irreducible octahedral representations
in the cluster can yield eigenmode solutions, though just as with
all the rest of the atoms in the crystal, the $u_{\alpha}(\ell)$
displacements associated with these other irreducible representations
will, via Eq. (\ref{5}), still participate in the $F_{1u}$ mode
oscillations.

With the eigenmodes associated with the $F_{1u}$ sector thus exhausting
the eigenspectrum, it will be these modes alone which will contribute to
the trace condition of Eq. (\ref{15}). And with the zeroes of
$S(\omega^{\prime 2})$ still obeying  Eq. (\ref{33}), for a defect with
$M^{\prime}=M$ in a crystal with central force-constants alone and
eigenmodes which then obey
\begin{equation}
\sum_{i=1}^{N} [\omega^{\prime
2}_i-\omega_i^2]=\sum_{i=1}^{N-1}[\omega^{\prime 2}_i
-\omega^{2}_{i}]+\omega^{\prime 2}_{\rm max} -\omega^{ 2}_{\rm
max}=\frac{2A^{\prime}_{xx}(0,0)}
{M}-\frac{2A_{xx}(0,0)}{M}~~,
\label{41}
\end{equation}
and in-band modes which can never be shifted up beyond the immediate
next zeroes of $S(\omega^{\prime 2})$ (typical curve $AX$ in Fig. (2)), 
we can conclude that no matter which central force-constant crystal
we may choose, there will definitely be a localized mode if
$A_{xx}^{\prime}(0,0)/A_{xx}(0,0)$ is greater than $3/2$, with this
condition thus being sufficient to guarantee localized modes in
nearest-neighbor central force-constant cubic crystals when there is no
change in mass.\cite{footnote9}

\section{Extension to more general situations}

To extend these results to other cases, we need to consider both
non-central force-constants and go beyond nearest neighbors. As regards
the issue of non-central force-constants, we note that within the
nearest-neighbor approximation a complete and exact general relation
for locating the perturbed crystal $F_{1u}$ modes has recently actually
been obtained.\cite{Mannheim2005} Specifically, it was found for the
body-centered cubic crystal (and thus by extension for the face-centered
cubic crystal as well since it is the reciprocal lattice of the
body-centered cubic and we work in the harmonic approximation where
momentum and position are treated equivalently)\cite{footnote10} that the
general determinantal condition $|1-G_0V|=0$ in the perturbed $F_{1u}$
mode sector can be written as $|1-G_0V|=\Delta^3=0$ where the
nearest-neighbor, arbitrary force-constant $\Delta$ is given as
\begin{equation}
\Delta=\frac{M^{\prime}}{M}\left\{\frac{A^{\prime}_{xx}(0,0)}{A_{xx}(0,0)}
\left[1-\rho(\omega^{\prime 2})
S(\omega^{\prime 2})\right]\left[1-\hat{R}\right]
-\mu\hat{R}\left[S(\omega^{\prime 2})\left(1-\frac{M}{M^{\prime}}\right)
+1\right]\right\}~~,
\label{42}
\end{equation}
expressed here in terms of the two quantities $\rho(\omega^{\prime 2})$
and $S(\omega^{\prime 2})$ which were given previously and two additional
quantities $\mu$ and $\hat{R}$ which are needed now. To
define these two additional quantities one needs to introduce the pure
crystal lattice Green's function combination 
\begin{eqnarray}
R&=&g_{xy}(\omega^{\prime};\ell=0,\ell^{\prime}=222)
+g_{xy}(\omega^{\prime};\ell=0,\ell^{\prime}=220)
\nonumber \\
&=&
\frac{1}{NM}\sum_{\vec{\bf k},j}\frac{\sigma_{x}^{*j}
(\vec{\bf k})\sigma_{y}^{j}(\vec{\bf k})}{[\omega_{j}^2(\vec{\bf
k})-\omega^{\prime 2}]}
\left[e^{i\vec{\bf k}\cdot\vec{\bf R}(222)}+
e^{i\vec{\bf k}\cdot\vec{\bf R}(220)}\right]~~, 
\label{43}
\end{eqnarray}
and in terms of the interatomic potential $\phi(r)$ between two pure
crystal nearest neighbors and the interatomic potential $\hat{\phi}(r)$
between the defect and a host crystal nearest neighbor, define pure and
impure lattice force-constants via 
\begin{eqnarray}
&&A_{\alpha\beta}(0,111)= \pmatrix{
\alpha+\beta&\alpha&\alpha\cr
\alpha&\alpha+\beta&\alpha\cr
\alpha&\alpha&\alpha+\beta\cr}~~,~~
A^{\prime}_{\alpha\beta}(0,111)= \pmatrix{
\hat{\alpha}+\hat{\beta}&\hat{\alpha}&\hat{\alpha}\cr
\hat{\alpha}&\hat{\alpha}+\hat{\beta}&\hat{\alpha}\cr
\hat{\alpha}&\hat{\alpha}&\hat{\alpha}+\hat{\beta}\cr
}~~,
\nonumber \\
&&A_{xx}(0,0)=-8(\alpha+\beta)~~,~~
A^{\prime}_{xx}(0,0)=-8(\hat{\alpha}+\hat{\beta})~~,
\label{44}
\end{eqnarray}
where 
\begin{eqnarray}
\alpha&=&-\frac{1}{3}\left(\phi^{\prime\prime}(r)
-\frac{\phi^{\prime}(r)}{r}\right)~~,~~\beta
=-\frac{\phi^{\prime}(r)}{r}~~,~~
\nonumber \\
\hat{\alpha}&=&-\frac{1}{3}\left(\hat{\phi}^{\prime\prime}(r)
-\frac{\hat{\phi}^{\prime}(r)}{r}\right)~~,~~\hat{\beta}
=-\frac{\hat{\phi}^{\prime}(r)}{r}
\label{45}
\end{eqnarray}
(as evaluated at the pure crystal nearest-neighbor equilibrium
separation $r$), with $\hat{R}$, $\mu$ and $\mu\hat{R}$ then being given
by 
\begin{eqnarray}
\hat{R}&=&
\frac{(\alpha+\beta)(\beta-\hat{\beta})(3\alpha-3\hat{\alpha}
+\beta-\hat{\beta})R}
{\alpha(\alpha-\hat{\alpha}+\beta-\hat{\beta})}~~,
\nonumber \\
\mu&=&
\frac{2(\alpha\hat{\beta}-\beta\hat{\alpha})^2}
{(\alpha+\beta)^2
(\beta-\hat{\beta})(3\alpha-3\hat{\alpha}+\beta-\hat{\beta})}~~,
\nonumber \\
\mu\hat{R}&=&
\frac{2(\alpha\hat{\beta}-\beta\hat{\alpha})^2R}
{\alpha(\alpha+\beta)
(\alpha-\hat{\alpha}+\beta-\hat{\beta})}~~.
\label{46}
\end{eqnarray}
Equation (\ref{42}) is not only a very compact relation (it only requires
knowledge of two pure crystal Green's functions combinations, viz.
$g_{xx}(\omega^{\prime};0,0)$ and $R$),\cite{footnote11} but it
additionally reduces to the  previous condition given in Eq. (\ref{39})
whenever $\mu\hat{R}$ is zero. Now while the quantity $\mu\hat{R}$ would
vanish when
$\beta=\hat{\beta}=0$, viz. the previously discussed  central
force-constant case, it would also vanish when $\alpha=\hat{\alpha}=0$,
viz. the isotropic force-constant case, and also whenever
$\hat{\alpha}/\alpha$ and $\hat{\beta}/\beta$ are equal, viz. in the
mixed case in which the fractional changes in the central and isotropic
components of the force-constants are equal to each other, with
$A^{\prime}_{xx}(0,0)/A_{xx}(0,0)=
(\hat{\alpha}+\hat{\beta})/(\alpha+\beta)$ being equal to
$\hat{\alpha}/\alpha =\hat{\beta}/\beta$ in all such cases. For all of
these cases then, the condition $A^{\prime}_{xx}(0,0)/A_{xx}(0,0) >3/2$
is sufficient to guarantee localized modes in nearest-neighbor crystals
when there is no change in mass.

To treat cases where the quantity $\mu$ is not zero is not as
straightforward, since unlike the
$g_{xx}(\omega^{\prime };0,0)$ Green's function, the $R$ combination 
cannot be reduced to a sum solely over pure crystal eigenfrequencies as
the polarization vectors cannot readily be eliminated from Eq.
(\ref{43}), with Eq. (\ref{42}) not obviously being reducible to an
expression which only involves the pure crystal density of states
$\nu(\omega^2)$. However, despite this, it is still possible to extract a
sufficiency condition for localized modes. Specifically, even though $R$
does involve the pure crystal polarization vectors, as can be seen from
Eq. (\ref{43}), its poles, and thus its asymptotes, are nonetheless
precisely the same as those possessed by
$g_{xx}(\omega^{\prime};0,0)$, with the asymptotes of
$\Delta$ then being none other than the asymptotes of $S(\omega^{\prime
2})$, i.e. none other than the ones exhibited in Fig. (2). To determine
what happens when we take $\mu$ to be non-zero then, we need to monitor
how the intercept structure displayed in Fig. (2) gets modified as we
slowly switch $\mu$ on.

\begin{figure}
\centerline{\epsfig{file=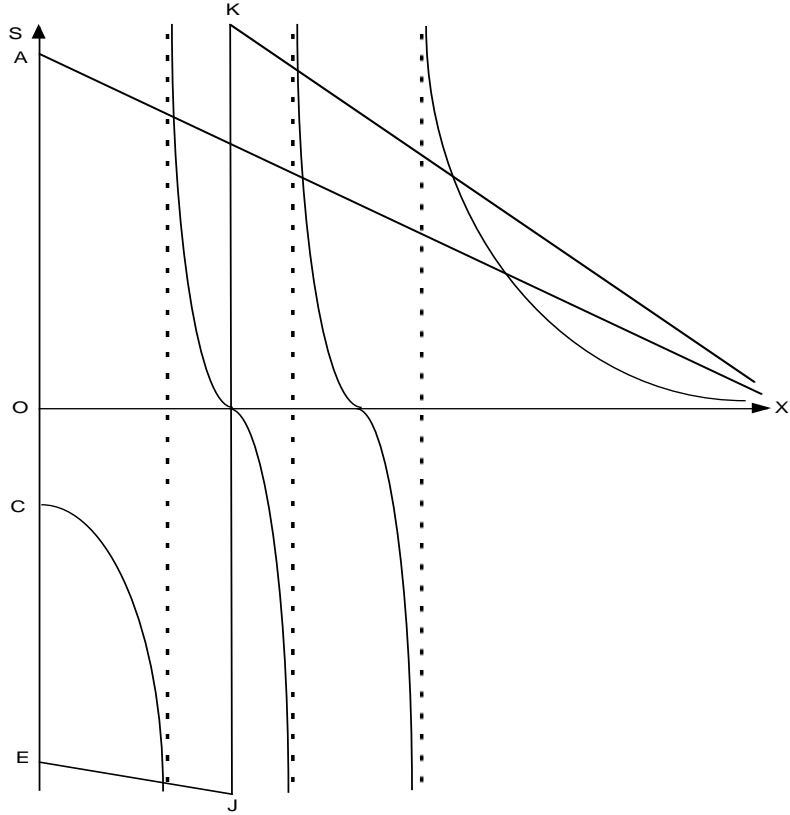,width=5.1in,height=5.1in}}
\medskip
\caption{Plot of $S(\omega^{\prime 2})$ versus $\omega^{\prime
2}$. The line $AX$ corresponds to the value of $1/\rho(\omega^{\prime 2})$
associated with a defect with $M^{\prime}=M$ and
$A_{xx}^{\prime}(0,0)>A_{xx}(0,0)$, and the
curve $EJKX$ corresponds to the $1/\rho(\omega^{\prime 2})$
associated with the same
$A_{xx}^{\prime}(0,0)>A_{xx}(0,0)$  but with $M^{\prime}$ slightly
larger than $M$.}
\label{Fig. (3)}
\end{figure} 

To explicitly see what happens when we switch $\mu$ on, it is instructive
to first return to the central force-constant case, and consider the
situation in which we start with some arbitrary $A^{\prime}_{xx}(0,0)$
which is bigger than $A_{xx}(0,0)$ and some initial $M^{\prime}$ which is
equal to $M$, and them slowly start to increase $M^{\prime}$. We thus
start with a $\rho(\omega^{\prime 2})$ which is equal to $(2\omega^{\prime
2}/\omega_{\rm max}^2)[1-A_{xx}(0,0)/A^{\prime}_{xx}(0,0)]$, and thus
with a $1/\rho(\omega^{\prime 2})$ which is everywhere positive and
infinite at $\omega^{\prime 2}=0$ (viz. curve $AX$ of Fig. (3)). When we
now allow $M^{\prime}$ to be just a little bit bigger than $M$,
$\rho(\omega^{\prime 2})$ will now take small negative values at the
smallest $\omega^{\prime 2}$, and then quickly revert back to being
positive again as $\omega^{\prime 2}$ is increased. Consequently,
$1/\rho(\omega^{\prime 2})$ will now take large negative values at the
smallest $\omega^{\prime 2}$, while also quickly reverting back to
being positive again as $\omega^{\prime 2}$ is increased (viz. curve
$EJKX$ of Fig. (3)). As can therefore be seen from Fig. (3), the net
effect of letting $M^{\prime}$ be just a little bit bigger than $M$ is
that rather than increasing, the frequency of the lowest mode is instead
decreased. Now since the trace condition of Eq. (\ref{15}) holds for both
of these two cases, we can write a trace condition for the difference
between the two cases, viz.
\begin{eqnarray}
\sum_{i=1}^{N} [\omega^{\prime
2}_i(M^{\prime}>M)-\omega_i^{\prime 2}(M^{\prime}=M)]=
\frac{A^{\prime}_{xx}(0,0)}{M^{\prime}}
-\frac{A^{\prime}_{xx}(0,0)}{M}~~,
\label{47}
\end{eqnarray}
with the net effect of a small difference between $M^{\prime}$ and $M$ on
the right-hand side of Eq. (\ref{47}) entailing a small net total shift on
the left-hand side, and thus a net shift in each individual eigenmode of
order $1/N$ of the shift on the right-hand side. Moreover, with the change
in the right-hand side of Eq. (\ref{47}) being continuous, the change on
the left-hand side must be continuous as well. Consequently, with the
transition from curve $AX$ to curve $EJKX$ needing to also be continuous,
the lowest impure crystal eigenfrequency must then have continuously
traversed the lowest lying of the pure crystal asymptotes on its way,
with the lowest lying pure crystal eigenfrequency then being an impure
crystal eigenfrequency at the point at which the asymptote is reached.
Since $S(\omega^{\prime 2})$ is infinite at any pure crystal
eigenfrequency, for such a pure crystal eigenfrequency to also be a mode
of the impure crystal, $1/\rho(\omega^{\prime 2})$ must equally be
infinite at the pure crystal eigenfrequency, and it indeed is in this
particular case since an increase in mass and an increase in
force-constant act oppositely in
$\rho(\omega^{\prime 2})$, to thereby allow $\rho(\omega^{\prime 2})$ to
indeed vanish at the lowest pure crystal eigenfrequency. 

An alternate way to modify the curve $AX$ is not to change the mass at
the defect site at all, but to keep $M^{\prime}$ fixed at $M$ and
to instead start to allow an $A^{\prime}_{xx}(0,0)$ which is already
bigger than $A_{xx}(0,0)$ to get even bigger. Such an increase will cause
all of the impure crystal eigenfrequencies of Fig. (\ref{3}) to increase,
but will never permit any of them to ever reach or traverse the next
immediate pure crystal asymptote since a $\rho(\omega^{\prime 2})=
(2\omega^{\prime 2}/\omega_{\rm
max}^2)[1-A_{xx}(0,0)/A^{\prime}_{xx}(0,0)$ which only  involves
force-constant changes can never vanish at any non-zero
$\omega^{\prime 2}$. We thus recognize two possible options as we start to
vary parameters, namely that an impure crystal eigenmode can cross a pure
crystal asymptote if $\rho(\omega^{\prime 2})$ has a zero there, or
cannot do so if $\rho(\omega^{\prime 2})$ has no zero. If we thus picture
the $N$ asymptotes in Fig. (3) as dividing the $\omega^{\prime 2}>0$
region into $N+1$ compartments ($N-1$ of which lie between pure crystal
eigenfrequencies, with the other two lying below the lowest pure crystal
eigenfrequency and above the largest one), we see that small changes in
parameters can cause the locations of eigenmodes to either move slightly
within any given compartment or to cross into an adjacent one, doing so
in either case without radically altering the value of the total in-band
difference $\sum_{i=1}^{N-1} [\omega^{\prime 2}_i-\omega_i^2]$ contained
on the left-hand side of Eq. (\ref{15}).

To see how this analysis pertains to non-central force-constant crystals
with non-zero $\mu$, we need to determine whether or not switching on
$\mu$ can cause modes to change compartments. To make such a
determination, we  note that according to Eq. (\ref{42}), in the event of
no mass change at the defect site, the eigenfrequencies associated with a
general force-constant change in a nearest-neighbor crystal are given as
the solutions to
\begin{eqnarray}
S(\omega^{\prime
2})&=&\frac{1}{\rho(\omega^{\prime
2})}\left[1-\frac{ A_{xx}(0,0)\mu\hat{R}}{A^{\prime}_{xx}(0,0)[1-\hat{R}]}
\right]
\nonumber \\
&=&\frac{\omega_{\rm
max}^2A_{xx}(0,0)}{2\omega^{\prime
2}[A^{\prime}_{xx}(0,0)-A_{xx}(0,0)]}
\left[\frac{A^{\prime}_{xx}(0,0)}{A_{xx}(0,0)}-\frac{
\mu\hat{R}}{[1-\hat{R}]}\right]~~.
\label{48}
\end{eqnarray}
Since the pure crystal Green's function $\hat{R}$ has the same set of
asymptotes as $S(\omega^{\prime 2})$, whenever $S(\omega^{\prime 2})$
diverges, $\hat{R}$ will do so also. However since $\hat{R}/(1-\hat{R})$
is finite at frequencies at which $\hat{R}$ is infinite, we see from Eq.
(\ref{48}) that in the force-constant change case no matter what value we
allow for $\mu$, the perturbed eigenmodes are unable to ever leave their
respective frequency compartments.

Despite this though, since $\hat{R}$ does have
asymptotes, it can be expected that $\hat{R}$ would change sign on its
way between adjacent asymptotes (though it could drop to some minimum
and then go back to the next asymptote without ever changing sign), and
thus we can anticipate that there will be some frequencies at which
$\hat{R}$ can take an assigned value equal to one. At such frequencies the
right-hand side of Eq. (\ref{48}) would then diverge, and since $\hat{R}$
(and thus $S(\omega^{\prime 2})$) is not itself diverging at those points,
such points would not be solutions to Eq. (\ref{48}), but would instead
be points between pure crystal eigenfrequencies at which the right-hand
side of Eq. (\ref{48}) would undergo a discontinuity. Now while such a
discontinuity cannot take the eigenmode out of its frequency compartment,
if the discontinuity is to have any effect at all, its only possible one
would be to move the eigenmode out of the $F_{1u}$ mode sector determinant
altogether. Specifically, unlike the case of a perturbation of a central
force-constant crystal where the $F_{1u}$ sector determinant accounts for
all of the perturbed crystal eigenmodes, once we introduce non-central
force-constants we have to consider the possibility that eigenmodes could
move into the determinantal blocks of $|1-G_0V|$ which are associated with
the other irreducible octahedral representations of the cluster as given
in Eq. (\ref{38}). However, recalling that all such eigenmodes are also
controlled by Eq. (\ref{20}), and recalling that all the pure lattice
Green's functions which appear in Eq. (\ref{20}) have the common
asymptote structure associated with Eq. (\ref{18}) and thus a compartment
structure identical to that of $S(\omega^{\prime 2})$ itself, we see that
even if the eigenmodes do migrate into different irreducible $O_h$
sectors, since the left-hand side of the trace condition of Eq.
(\ref{15}) is a sum over all of the modes of the crystal and not just
over those associated with the $F_{1u}$ sector alone (a sum that thus
includes the entire cluster of Eq. (\ref{38}) anyway), and since we are
only making a small change on the right-hand side of Eq. (\ref{15}), when
the eigenmodes do move to other irreducible representations, they can
still only move to equivalent compartments with the same frequency
ranges as the ones they started in. As such then, independent of which
specific sector the modes actually find themselves in after crossing
any $\hat{R}=1$ discontinuity, they could only cross into the same
frequency compartments they were already in.

However, once the eigenmodes have moved
out of the $F_{1u}$ sector, they would then be in some other
irreducible $O_h$ representation where the $F_{1u}$ mode Eq.
(\ref{48}) would not apply. Hence in these other representations
additional changes in $\mu$ (beyond the ones needed to first bring the
force-constant dependent $\hat{R}$ of Eq. (\ref{46}) to one) could perhaps
then move the eigenmodes into adjacent compartments. However, for that to
happen the eigenmodes would have to cross the pure crystal asymptotes,
with those particular eigenmodes then needing to be common eigenmodes of
both the pure and impure crystals at the requisite  value of $\mu$.
However, in order for an eigenmode to be an eigenmode at all, the
eigenmode would need to be at a zero of the full $|1-G_0V|$, i.e. it would
not only need to be at a zero of the determinantal block of some
particular irreducible sector, but also to not give an infinite value to
any of the other blocks which multiply the given determinantal block of
interest in the full $|1-G_0V|$ of the full crystal. However, inspection
of the $F_{1u}$ mode $\Delta$ of Eq. (\ref{42}) shows that it is
actually divergent at a pure crystal eigenmode (where it diverges as 
$S(\omega^{\prime 2})\hat{R}$), and thus once there is no asymptote
crossing in the $F_{1u}$ mode sector, there cannot be any in any other
irreducible representation either. Thus regardless of whether or not the
$\hat{R}=1$ discontinuities take the impure crystal eigenmodes out of the
$F_{1u}$ sector, the modes cannot leave the frequency compartments they
were already in when $\mu$ was zero. Hence if a mode already was in the
compartment which lies beyond the pure crystal maximum eigenfrequency
when the parameter $\mu$ was equal to zero, the mode will remain beyond
the band no matter how $\mu$ might then vary. Moreover the in-band modes
will remain in which ever compartments they had been in when $\mu$ was
zero. 

We thus recognize two specific consequences of switching $\mu$ on,
namely that modes have to stay within their respective $\mu=0$
compartments, but that they are no longer prevented from traversing the
zeroes of $S(\omega^{\prime 2})$. Thus suppose we start off with a central
force-constant crystal with $\mu=0$ and $M^{\prime}=M$, and with
$A^{\prime}_{xx}(0,0)/A_{xx}(0,0)=(\hat{\alpha}+\hat{\beta})/(\alpha+\beta)
=\hat{\alpha}/\alpha=\hat{\beta}/\beta$ conveniently set right at the
threshold value of $3/2$, and then slowly switch $\mu$ on. The in-band
modes could now move up or down, moving to no lower than the immediately
previous pure crystal eigenmodes (net shift of the in-band mode
$\sum_{i=1}^{N-1} [\omega^{\prime 2}_i-\omega_i^2]$ of down to 
$-\omega_{\rm max}^2$), or moving upwards possibly as far as the next
immediate pure crystal asymptotes (net shift of up to $\omega_{\rm
max}^2$).\cite{footnote12} Then regardless of what value $\mu$ may
actually take and regardless of which particular irreducible
representation any specific eigenmode may actually lie in, Eq. (\ref{41})
will nonetheless still hold when there is no mass change since the trace
is taken over the entire set of irreducible $O_h$ representations and not
just over the
$F_{1u}$ mode. On recalling that
$\omega_{\rm max}^2=2A_{xx}(0,0)/M$ for a nearest-neighbor crystal, we
see that the necessary condition that $\omega^{\prime 2}_{\rm max}-
\omega_{\rm max}^2$ be positive is given by having the in-band modes move
down as far as they possibly can, to thereby require
$A^{\prime}_{xx}(0,0)$ to be positive, while the sufficient condition is
given by having the in-band modes move up by as far as they possibly can,
viz. 
\begin{equation}
\omega^{\prime 2}_{\rm max}- \omega_{\rm max}^2 \geq 
\frac{2A^{\prime}_{xx}(0,0)}{M}
-\frac{4A_{xx}(0,0)}{M}~~, 
\label{49}
\end{equation}
with the condition that $A_{xx}^{\prime}(0,0)/A_{xx}(0,0)$ be greater than
two thus being sufficient to guarantee the generation of localized modes 
by force-constant changes alone in an arbitrary nearest-neighbor cubic
crystal.

To proceed beyond nearest neighbors is also not straightforward as 
no exact analog of Eqs. (\ref{39}) and (\ref{42}) is currently known.
(For the case of second-nearest neighbors, the second-neighbor
generalization of the irreducible decomposition of Eq. (\ref{38}) 
has been used to bring the problem to a reasonably manageable
form.\cite{Lakatos1968}) However, in the event of non-nearest
neighbor force-constants, even though the cluster required for the matrix
$V_{\alpha\beta}(\ell^{\prime},\ell^{\prime\prime})$ in Eq. (\ref{20})
would be much larger, and even though more pure lattice Green's functions
would be required, nonetheless every single one of these Green's
functions would still have exactly the same asymptote and compartment
structure as $S(\omega^{\prime 2})$. Thus if we start at the eigenmodes of
a general nearest-neighbor crystal with non-zero $\mu$ and
$M^{\prime}=M$ and slowly start to increase the strength of the
non-nearest neighbor force-constants (central or otherwise), we would
again be constrained by the same set of pure crystal compartments in
exactly the same way as before (save that there would be yet more
irreducible representations of $O_h$ to migrate to -- with all of them
also contributing to the left-hand side of the trace condition), to again
allow us to infer that even in the most general possible case imaginable
(arbitrary central and non-central force-constants and arbitrary number
of participating neighbors), in the absence of any change in mass, the
condition $A_{xx}^{\prime}(0,0)/A_{xx}(0,0)>2$ would still be sufficient
to guarantee localized modes.\cite{footnote13}

\begin{acknowledgments}
Both of the authors would like to thank Dr. E. E. Alp for the kind
hospitality of the Advanced Photon Source at Argonne National Laboratory
where this work was performed.
\end{acknowledgments}


\begin{thebibliography}{}
\bibitem{footnote1} The force-constants yielded by potentials for which
$\phi^{\prime}(r)$ is zero when the atoms are in their equilibrium
positions are referred to as being central, while those associated with
potentials which obey $\phi^{\prime}(r)/r=\phi^{\prime\prime}(r)$ at
equilibrium are referred to as being isotropic.
\bibitem{footnote3} For any choice of the $M^{\prime}/M$ ratio, the curve
$1/\rho(\omega^{\prime 2})=M^{\prime}/(M-M^{\prime})$ is parallel to the
$OX$ axis in Fig. (1), lying above the line $OX$ when $M>M^{\prime}$ and
below the line
$CD$ when
$M^{\prime}>M$, being at $\pm \infty$ when $(M-M^{\prime})=0\pm$.
\bibitem{footnote5} For the face-centered cubic crystal for instance the
normalized four-dimensional basis of displacements for the $4F_{1u}$ mode
$(\alpha_0,\alpha_1,\alpha_2,\alpha_3)$ is given (in face-centered cubic
crystal site notation) by
$\alpha_0=u_x(0,0,0)$,
$\alpha_1=(1/2\sqrt{2})[u_x(1,1,0)+u_x(\bar{1},\bar{1},0)+u_x(1,0,1)
+u_x(\bar{1},0,\bar{1})+u_x(1,\bar{1},0)+u_x(\bar{1},1,0)
+u_x(\bar{1},0,1)+u_x(1,0,\bar{1})]$,
$\alpha_2=(1/2\sqrt{2})[u_y(1,1,0)+u_y(\bar{1},\bar{1},0)+u_z(1,0,1)
+u_z(\bar{1},0,\bar{1})-u_y(1,\bar{1},0)-u_y(\bar{1},1,0)
-u_z(\bar{1},0,1)-u_z(1,0,\bar{1})]$,
$\alpha_3=(1/2)[u_x(0,1,1)+u_x(0,\bar{1},\bar{1})+u_x(0,1,\bar{1})
+u_x(0,\bar{1},1)]$. Degenerate with this basis are two others, one
based on $u_y(0,0,0)$ and the other on $u_z(0,0,0)$.
\bibitem{footnote6} On the $AGHX$ curve the loss of an intercept outside
the band is compensated for by having an additional intercept in the
region where the $1/\rho(\omega^{\prime 2})$ divergence is located, with
all modes below the line $GH$ being shifted upwards, and all modes above
the line $GH$ being shifted downwards.
\bibitem{footnote7} After first intercepting the $S(\omega^{\prime 2})$
curve $N$ times in the positive region before reaching the point where
it diverges, on reaching this point the function
$1/\rho(\omega^{\prime 2})$ then crosses into the negative region with
no further intercepts then being encountered.
\bibitem{footnote8} On the $EJKX$ curve the absence of any intercept at
all in the region where the $1/\rho(\omega^{\prime 2})$ divergence is
located is compensated for by having an additional intercept outside the
band, with all modes below the line $JK$ being shifted downwards, and all
modes above the line $JK$ being shifted upwards.
\bibitem{footnote10} While not sought here, one can anticipate a
qualitatively similar form for the determinant $\Delta$ in the simple
cubic case.
\bibitem{footnote12} The presence of any analogs of the zeroes of
$S(\omega^{\prime 2})$ in the other irreducible $O_h$ sectors could
prevent the eigenmodes from actually being able to go all the way to the
next adjacent asymptotes.
\bibitem{footnote13} To be slightly more precise, in relating
$\omega_{\rm max}^2$ to $2A_{xx}(0,0)/M$ in the non-nearest neighbor
case, one should also take into account the factor $Z$ as given in
Eq. (\ref{31}), though we neglect this effect here.
\end{thebibliography}
\end{document}